\documentclass[slac_one]{revtex4}
\usepackage{graphicx}
\usepackage{fancyhdr}
\usepackage{amsbsy}
\usepackage{amsmath}
\usepackage{amsfonts}
\usepackage{amssymb}
\begin{document}

\title{Minimal Flavor Violation as an alternative to R-parity}

\author{Christopher Smith}
\affiliation{Institut f\"ur Theoretische Physik, Universit\"at Bern, Sidlerstrasse 5, CH-3012 Bern, Switzerland}

\begin{abstract}
A peculiar aspect of the MSSM, the simplest supersymmetric extension of the Standard Model, is that it is usually defined including an ad hoc symmetry, R-parity, whose sole purpose is to forbid rapid proton decay. This symmetry deeply alters the phenomenology of the MSSM, and renders the experimental search strategies quite involved. Besides, the MSSM suffers from a number of flavor puzzles. Generically, the superparticle contributions to Flavor Changing Neutral Currents (FCNC) are too large compared to experiment, both in the quark and lepton sectors. The Minimal Flavor Violation (MFV) hypothesis aims at suppressing these contributions, and when enforced as a symmetry principle, achieves this in a very natural and systematic way. In this talk, it will be shown that imposing MFV is not only able to suppress supersymmetric contributions to FCNC, it also naturally explains the apparent stability of the proton. As a result, R-parity can be avoided altogether, motivating the search for supersymmetry through simpler channels, like for example single stop resonant production, whose strength is predicted by MFV.
\end{abstract}

\maketitle

\thispagestyle{fancy}

\section{INTRODUCTION}

The Minimal Supersymmetric extension of the Standard Model (MSSM) introduces many additional flavored particles -- one scalar squark (slepton) superpartner for each of the chiral quark (lepton) fields. If light, their presence forces us to fine-tune the parameters of the MSSM in order to satisfy all experimental constraints. This is the so-called MSSM flavor puzzle. For instance, the extensive and precise data from B and K physics on Flavor-Changing Neutral Currents (FCNC) requires limiting squark-induced flavor mixings, including the possible additional CP-violating phases occurring there. Similarly, the non-observation of Lepton Flavor Violating (LFV) transitions like $\mu\rightarrow e\gamma$ restricts the acceptable level of flavor-mixing in the slepton sector. In these contexts, the Minimal Flavor Violation (MFV) hypothesis~\cite{HallR90,DambrosioGIS02} aims at restoring naturalness, by automatically limiting the maximal amount of flavor mixings in both the squark and slepton sectors (for phenomenological applications in the MSSM, see e.g.~\cite{Pheno,RGE}).

There is yet another, more puzzling consequence of the presence of the squarks and sleptons. These scalar particles can be used to write down renormalizable Yukawa-type interactions violating baryon ($\mathcal{B}$) and lepton ($\mathcal{L}$) numbers, which then drive proton decay at tree-level. This is a severe problem given the current experimental bounds on the proton lifetime, well-beyond 10$^{30}$ years for most decay modes. These bounds are actually so tight that the MSSM is usually equipped with an ad-hoc symmetry, R-parity~\cite{FarrarF78}, which simply forbids all the $\Delta\mathcal{B}=1$ and $\Delta\mathcal{L}=1$ interactions. In this talk, we will present a different approach, in which all these couplings are not discarded outright. Instead, since these $\Delta\mathcal{B}=1$ and $\Delta\mathcal{L}=1$ interactions are also necessarily flavor-changing, proton stability will be considered as a manifestation of the MSSM flavor puzzle. Then, MFV will be called in to constrain all the R-parity violating couplings, by using exactly the same strategy as for FCNC. That MFV is a viable alternative to R-parity will finally emerge naturally~\cite{NikolidakisS07}, and will force a reassessment of possible supersymmetric signals at colliders.

\section{MFV AND THE ORIGIN OF THE FLAVOR STRUCTURES}

The formulation of MFV is detailed in Refs.~\cite{DambrosioGIS02,RGE,NikolidakisS07}, and is only sketched here. To
start with, one observes that the three generations of (s)quarks and
(s)leptons have identical gauge interactions. In other words, the gauge sector of the MSSM exhibits a global $U(3)^{5}$ flavor-symmetry, one $U(3)$ for each of the quark and lepton superfields~\cite{ChivukulaG87}
\begin{equation}
G_F\equiv U(3)^5=G_q\times G_{\ell}\times U(1)^5,\;\text{with}\;G_q\equiv SU(3)_Q\times SU(3)_U\times SU(3)_D,\;G_{\ell}\equiv SU(3)_L\times SU(3)_E\;.
\end{equation}
Of course, this symmetry is broken in many sectors of the MSSM --
specifically, by the R-parity conserving and violating superpotential and soft SUSY breaking couplings. However, at present, the only observed breakings of $G_{F}$ are the non-equality of quark and lepton masses, and their mixings. All these entirely derive from the usual Yukawa couplings $\mathbf{Y}_{u,d,e}$, introduced in the MSSM through the superpotential $\mathcal{W}$ (up to neutrino masses and mixings, discussed later)
\begin{equation}
\mathcal{W}\supset U\mathbf{Y}_{u}QH_{u}-D\mathbf{Y}_{d}QH_{d}-E\mathbf{Y}_{e}LH_{d}\;.\label{W}
\end{equation}
Therefore, the starting point of MFV is to assume that these Yukawa
interactions actually represent the maximal flavor-symmetry breakings allowed in the full MSSM.

This does not mean that the other MSSM sectors where $G_{F}$ is broken are ruled out, but forces all the parameters occurring there to be related to the Yukawa couplings. To systematically investigate these relationships, MFV is formulated as a symmetry principle: The Yukawa interactions in Eq.(\ref{W}) are formally $G_{F}$-invariant if the Yukawa couplings are promoted to spurion fields, i.e. are made to transform under $G_{F}$ just such as to balance the transformations of the quark and lepton superfields:
\begin{equation}
\mathbf{Y}_{u}\sim\left(  \bar{3},3,1\right)  _{G_{q}},\;\;\mathbf{Y}_{d}\sim\left(  \bar{3},1,3\right)  _{G_{q}},\;\;\mathbf{Y}_{e}\sim\left(  \bar{3},3\right)  _{G_{\ell}}\;.\label{Yukawas}
\end{equation}
These Yukawa spurions are then used to parametrize all the flavor-breaking couplings of the MSSM as formally $G_{F}$-invariant. Once this step is complete, $\mathbf{Y}_{u}$, $\mathbf{Y}_{d}$ and $\mathbf{Y}_{e}$ are frozen back to their physical values, thereby obtaining predictions for the strength of these couplings under MFV.

Having reconstructed all the flavor-breaking couplings in terms of the Yukawa couplings, they all become non-generic. They exhibit highly hierarchical structures directly inherited from those of the quark and lepton masses, as well as of the CKM matrix. In this way, the fine-tunings required to satisfy the experimental constraints on FCNC and LFV transitions are to a large extent automatically achieved, and are just as natural (or unnatural) as the Yukawa couplings themselves. In this sense, MFV does not explain the origin of the flavor structures, but reduces them to the minimal irreducible set of elementary flavor-breaking structures shown in Eq.(\ref{Yukawas}).

\boldmath
\section{THE $\Delta\mathcal{B}=1$ and $\Delta\mathcal{L}=1$ COUPLINGS ARE INTRINSICALLY DIFFERENT!}
\unboldmath

The gauge quantum numbers of the quark and lepton superfields are such that possible renormalizable $\Delta\mathcal{B}=1$ and $\Delta\mathcal{L}=1$ couplings have intrinsically different properties under the flavor group $G_{F}$. Indeed, while it is immediately possible to parametrize the former, $\boldsymbol{\lambda}^{\prime\prime IJK}U^I D^J D^K$, in terms of Yukawa spurions, for example as
\begin{equation}
\boldsymbol{\lambda}^{\prime\prime IJK}=\varepsilon^{LJK}(\mathbf{Y}_{u}\mathbf{Y}_{d}^{\dagger})^{IL},\;\varepsilon^{LMN}\mathbf{Y}_{u}^{IL}\mathbf{Y}_{d}^{JM}\mathbf{Y}_{d}^{KN},\;...\label{Eq1}
\end{equation}
the latter are forbidden as long as $m_{\nu}=0$. In other words, the charged lepton Yukawa spurion $\mathbf{Y}_{e}$ alone does not permit to write the $\boldsymbol{\mu}^{\prime I}L^{I}H_{d}$, $\boldsymbol{\lambda}^{IJK}L^{I}L^{J}E^{K}$ or $\boldsymbol{\lambda}^{\prime IJK}L^{I}Q^{J}D^{K}$ couplings as $G_{\ell}$-invariants.

To account for the tiny neutrino masses without introducing unnaturally small numbers, we supplement the MSSM with a seesaw mechanism of type I, by adding right-handed neutrino terms $\frac{1}{2}N^{T}\mathbf{M}_{R}N+N\mathbf{Y}_{\nu}LH_{u}$ to the superpotential~\cite{Seesaw}. Though this formally extends the flavor symmetry group to $G_{F}\times U(3)_N$, the heavy $N$ fields never occur at low-energy and only the spurion combinations which are singlets under $U(3)_N$ are relevant. Expanding in the inverse right-handed neutrino mass matrix $\mathbf{M}_{R}$, these are
\begin{equation}
\mathbf{Y}_{\nu}^{\dagger}\mathbf{Y}_{\nu},\;\mathbf{Y}_{\nu}^{T}%
(\mathbf{M}_{R}^{-1})\mathbf{Y}_{\nu},\;\;\;\mathbf{Y}_{\nu}^{T}(\mathbf{M}_{R}^{-1})(\mathbf{M}_{R}^{-1})^{\ast}\mathbf{Y}_{\nu}^{\ast},\;...\label{Eq5}
\end{equation}
The unsuppressed spurion, $\mathbf{Y}_{\nu}^{\dagger}\mathbf{Y}_{\nu}$, which tunes LFV effects as in the usual supersymmetric seesaw~\cite{BorzumatiM86}, transforms as $(8,1)$ under $G_{\ell}$, and is thus of no help to write the $\boldsymbol{\mu}^{\prime}$, $\boldsymbol{\lambda}$ or $\boldsymbol{\lambda}^{\prime}$ couplings as $G_{\ell}$-invariants. The only way to get flavor-symmetric operators is to use the neutrino mass spurion, $\Upsilon_{\nu}\equiv v_{u}\mathbf{Y}_{\nu}^{T}(\mathbf{M}_{R}^{-1})\mathbf{Y}_{\nu}\sim\mathcal{O}(m_{\nu}/v_{u})$, which transforms as $(\bar{6},1)$. For example, the $\boldsymbol{\mu}^{\prime}$, $\boldsymbol{\lambda}$ or $\boldsymbol{\lambda}^{\prime}$ couplings can be written as
\begin{align}
\boldsymbol{\mu}^{\prime I} & =\mu\varepsilon^{ILM}(\Upsilon_{\nu}^{\dagger})^{LM},\;...\label{Eq2}\\
\boldsymbol{\lambda} ^{IJK}  & =\varepsilon^{ILM}(\Upsilon_{\nu}^{\dagger})^{LM}\mathbf{Y}_{e}^{KJ},\;...\\
\boldsymbol{\lambda} ^{\prime IJK}  & =\varepsilon^{ILM}(\Upsilon_{\nu}^{\dagger})^{LM}\mathbf{Y}_{d}^{KJ},\;...
\end{align}
In fact, since the \={6} is symmetric, while $\varepsilon$-tensors are antisymmetric, additional $\mathbf{Y}_{e}$ insertions are needed to get non-vanishing contributions, e.g. $\varepsilon^{ILM}(\Upsilon_{\nu}^{\dagger})^{LM}\rightarrow\varepsilon^{ILM}(\mathbf{Y}_{e}^{\dagger}\mathbf{Y}_{e}\Upsilon_{\nu}^{\dagger})^{LM}$. We thus arrive at the conclusion that all $\Delta\mathcal{L}=1$ couplings are suppressed by the tiny neutrino masses as well as by the small charged-lepton masses.

\section{MFV CAN EXPLAIN THE VERY LONG PROTON LIFETIME}

To ensure that MFV is sufficient to prevent a too fast proton decay, we must construct all the possible $G_q\times G_{\ell}$-invariant operators, and check that all experimental bounds~\cite{Barbier04} are passed. We will not go through this analysis here (see Ref.~\cite{NikolidakisS07}), but rather take a simple illustrative example: the s-channel decay process shown in Fig.\ref{Fig1}.$a$, whose corresponding experimental bound is 
\begin{equation}
\left|\boldsymbol{\lambda}_{JMI}^{\prime}\boldsymbol{\lambda}_{11I}^{\prime\prime\ast}\right|  < 10^{-26}(m[\tilde{d}_{R}^{I}]/300\,GeV)^2\;,\label{Eq5b} 
\end{equation}
for $I,J=1,2,3$, $M=1,2$. Further, only the following three representative MFV operators for the $\boldsymbol{\lambda}^{\prime}$ and $\boldsymbol{\lambda}^{\prime\prime}$ couplings are kept (with $a_i\sim \mathcal{O}(1)$):
\begin{align}
\boldsymbol{\lambda}^{\prime IJK} & =a_{0}\varepsilon^{ILM}(\mathbf{Y}_{e}^{\dagger}\mathbf{Y}_{e}\Upsilon_{\nu}^{\dagger})^{LM}(\mathbf{Y}_{d}\mathbf{Y}_{u}^{\dagger}\mathbf{Y}_{u})^{KJ},\;\;\label{Eq6a}\\
\boldsymbol{\lambda}^{\prime\prime IJK} & =a_{1}\varepsilon^{LJK}(\mathbf{Y}_{u}\mathbf{Y}_{d}^{\dagger})^{IL}+a_{2}\varepsilon^{LMN}\mathbf{Y}_{u}^{IL}\mathbf{Y}_{d}^{JM}\mathbf{Y}_{d}^{KN}\;.\label{Eq6b}
\end{align}
Freezing the spurions to their physical values, the MFV prediction for the combination of couplings of Fig.\ref{Fig1}.$a$ reads
\begin{align}
\left| \boldsymbol{\lambda}_{JMI}^{\prime} \boldsymbol{\lambda}_{11I}^{\prime\prime\ast}\right| & \approx\frac{\sqrt{\Delta m_{atm}^{2}}}{v_{u}}\frac{m_{\tau}^{2}}{v_{d}^{2}}\lambda^{3}\frac{m_{b}m_{t}^{2}m_{u}}{v_{d}v_{u}^{3}}\left(  a_{0}a_{1}\frac{m_{s}}{v_{d}}+a_{0}a_{2}\frac{m_{d}m_{b}}{v_{d}^{2}}\right) \\ & \approx a_{0}a_{1}10^{-28}\tan^{4}\beta+a_{0}a_{2}10^{-31}\tan^{5}\beta\;\label{Eq3}
\end{align}
(assuming normal hierarchy, with $m_{\nu}^{lightest}=0$). Several suppression factors are immediately apparent. Besides the proportionality to the neutrino mass (actually, to the mass-difference), there are several light-quark mass factors, as well as CKM matrix elements, which together account for about half of the overall suppression. They all arise from the antisymmetry of the
$\varepsilon$-tensors, which forces the amplitude to be sensitive to the three generations.

These two mechanisms --neutrino mass factor \& $\varepsilon$-tensor
antisymmetry-- are sufficient to suppress proton decay below the current experimental limits. All the bounds are easily satisfied for small to moderate $\tan\beta$. For larger $\tan\beta$ values, several weaker suppression mechanisms could be at play -- heavier lightest neutrino, GIM-like interferences, restricting the broken flavor $U(1)$'s, reducing the MFV coefficients $a_{i}$,..., see Ref.~\cite{NikolidakisS07} for more details.

\section{MFV PREDICTIONS FOR R-PARITY VIOLATING PROCESSES}

MFV is a viable alternative to R-parity, and can give rise to very distinctive experimental signatures. Proton decay is an obvious target, since it can be quite close to the current experimental bound, but replacing R-parity by MFV can have profound consequences also in other sectors, in particular for SUSY searches at the LHC.
\begin{figure}[t]
\centering  
\includegraphics[width=150mm]{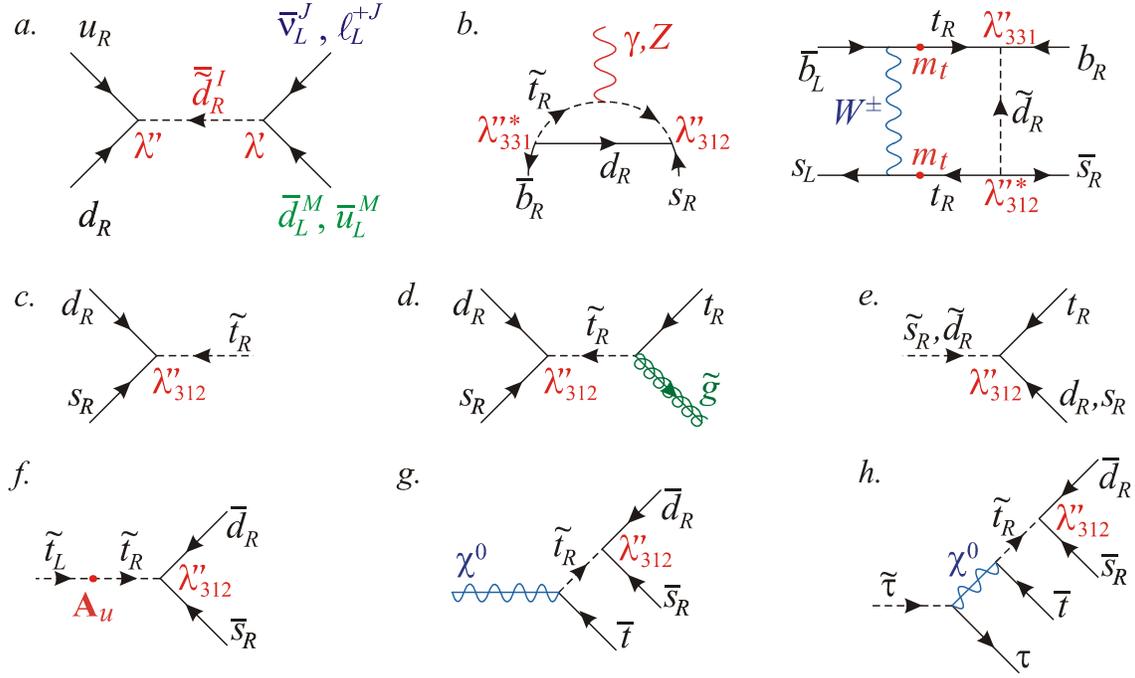}
\caption{$a$) Example of a tree-level, s-channel proton decay mechanism. $b$) Example of corrections to FCNC processes tuned by the $\Delta\mathcal{B}=1$ couplings. $c$--$h$) Possible consequences of $\Delta\mathcal{B}=1$ couplings with MFV strength: $c$) single stop resonant production~\cite{Stop} and $d$) its associated single gluino production~\cite{StopGluino}, $e$) top production from $\tilde{d}_R$ and $\tilde{s}_R$ decays~\cite{TopProd}, and $f$--$h$) possible $\tilde{t}_{L}$, $\chi^{0}$ and $\tilde{\tau}$ LSP decay cascades~\cite{LSP}.}%
\label{Fig1}
\end{figure}

All possible signals are driven by the $\Delta\mathcal{B}=1$ couplings, $\varepsilon^{abc}\boldsymbol{\lambda}_{IJK}^{\prime\prime}U_{a}^{I}D_{b}^{J}D_{c}^{K}$, since those violating $\mathcal{L}$ are proportional to neutrino masses, and thus tiny ($\boldsymbol{\mu}^{\prime}/\mu$, $\boldsymbol{\lambda}$ and $\boldsymbol{\lambda}^{\prime}<10^{-13}$). MFV predicts a very strong hierarchy for $\boldsymbol{\lambda}^{\prime\prime}$:
\begin{equation}
\boldsymbol{\lambda}_{312}^{\prime\prime}\sim10^{-1}\gg\boldsymbol{\lambda}_{331,323,212}^{\prime\prime}\sim10^{-4}>\boldsymbol{\lambda}_{IJK\neq 312,331,323,212}^{\prime\prime}\sim10^{-5}\;\text{or smaller}\;,\label{Eq4}
\end{equation}
so that only the $\tilde{t}_{R}s_{R}d_{R}$, $t_{R}\tilde{s}_{R}d_{R}$ and $t_{R}s_{R}\tilde{d}_{R}$ couplings are significant. This immediately implies that the impacts on FCNC are negligible, since at least two couplings with different flavor indices are needed. Specifically, the amplitudes for the diagrams shown in Fig.\ref{Fig1}.$b$ scale as~\cite{Barbier04,RPVFCNC}:
\begin{equation}
b\rightarrow s(d):|\boldsymbol{\lambda}_{312}^{\prime\prime\ast}%
\boldsymbol{\lambda} _{331(323)}^{\prime\prime}|\lesssim10^{-4}%
,\;\;\;s\rightarrow d:|\boldsymbol{\lambda} _{323}^{\prime\prime
\ast}\boldsymbol{\lambda}_{331}^{\prime\prime}|\lesssim10^{-7}\;,
\end{equation}
to be compared with $|V_{tb}^{\ast}V_{ts}|\sim10^{-2}$, $|V_{tb}^{\ast}V_{td}|\sim10^{-3}$ and $|V_{ts}^{\ast}V_{td}|\sim10^{-4}$ for the Standard Model contributions. At colliders, the situation is much more favorable. The most obvious processes induced by $\boldsymbol{\lambda}_{312}^{\prime\prime}$ are shown in Fig.\ref{Fig1}.$c$--$h$. Most of these signatures depend strongly on the precise mass spectrum of the MSSM, and have already been analyzed in details~\cite{Barbier04,General,LSP,Stop,StopGluino,TopProd}, though not within the MFV context yet. For example, whether the LSP decays in the detector or not depends on its nature~\cite{LSP}, since it may need to go through a lengthy cascade before reaching the $\boldsymbol{\lambda}_{312}^{\prime\prime}$ couplings to finally decay (see Fig.\ref{Fig1}.$f$--$h$).

\section{CONCLUSION}

The MFV hypothesis accounts simultaneously for the non-observation of New Physics effects in flavor transitions, and for the tremendously long proton lifetime. The latter is then understood as a direct consequence of the smallness of the neutrino masses and of the large hierarchies among quark masses and within the CKM matrix. Even if the origin of the flavor structures is left aside by MFV, naturalness is restored for all the flavor-dependent couplings of the MSSM. In particular, it now appears rather unnatural to allow for $\mathcal{O}(1)$ R-parity violating couplings, since such values correspond to breakings of the flavor-symmetry group that are many orders of magnitude larger than those due to the Yukawa couplings.

In view of these successes of MFV, whether R-parity should be imposed or not needs to be reassessed. Indeed, the proton decay puzzle is generally thought to be so serious that one is ready to accept the thorough alteration of the MSSM phenomenology induced by R-parity -- think in particular of the typical, but not experimentally friendly, topologies of SUSY events at the LHC. Once proton decay is taken care of by MFV, the motivations for nevertheless imposing R-parity have to be found elsewhere. In this respect, a few of the relevant issues for or against R-parity are: (1) Some dimension-five operators can induce proton decay even if R-parity conserving~\cite{IbanezR91}. On the other hand, they are again very suppressed by MFV. (2) In Grand Unified Theories, R-parity is often built in, or nevertheless required because their smaller flavor-symmetry group~\cite{MFVSU5} impeaches the MFV suppression to be as powerful. Still, it is fair to say that building a viable GUT with low-energy MFV has, up to now, not been fully investigated. (3) Without R-parity, the LSP of the MSSM is no longer a viable dark matter candidate. Remember though that there must be physics beyond the MSSM, since it is lacking a dynamical SUSY-breaking mechanism. (4) Having at hand several $\mathcal{B}$-violating couplings, which may also violate CP, could open new paths towards baryogenesis.

In conclusion, since the viability of the MSSM is no longer in the balance, whether R-parity is present in Nature or not is much less clear, and simple tree-level resonant productions of supersymmetric particles at the LHC may be just around the corner.

\begin{acknowledgments}

Work supported by the Swiss National Funds and by the EU RTN-CT-2006-035482 (Flavianet).

\end{acknowledgments}

\end{document}